\documentclass[12pt,a4paper]{article}

\usepackage{amsmath}
\usepackage{graphicx}
\usepackage{times}
\usepackage{cite}
\usepackage{ulem}
\begin{document}
\begin{titlepage}
\title{Detecting quantum fluctuations of multiplicity}
\author{ S.M. Troshin, N.E. Tyurin\\[1ex]
\small  \it NRC ``Kurchatov Institute''--IHEP\\
\small  \it Protvino, 142281, Russian Federation,\\
\small Sergey.Troshin@ihep.ru
}
\normalsize
\date{}
\maketitle

\begin{abstract}
	Transition to the reflective scattering mode   results in the increasing role of the multiplicity fluctuations of quantum origin and its asymptotic dominance. We  note here the feasibility to experimentally detect presence of quantum fluctuations of multiplicity   at finite energies. 
\end{abstract}
\end{titlepage}
\setcounter{page}{2}
\section*{Introduction}
Collision of  two quantum systems like hadrons evolves through the  transient state and leads to formation of two- or multiparticle final states. Quantum fluctuations of the initial systems and/or transient state result in  variation of the final state multiplicity. 

Fluctuations of the experimentally unmanageable initial impact parameter  should be considered separately. Use of the impact parameter representation provides a geometric, semiclassical picture of hadron interactions. At very high energies, this parameter in hadron scattering becomes a classical quantity and is approximately conserved, its commutator with Hamiltonian is vanishing \cite{webb} and unitarity equation for the scattering amplitude in the impact parameter representation is diagonalized at high energies.  It implies, in particular, a partial classicalization of this type of  multiplicity fluctuations.

Thus, it should be envisaged   the two different sources of the multiplicity fluctuations
in hadron production at available energies:  one  is related to  variations of the collision impact parameter value  and another  one should be associated with the quantum fluctuations of multiplicity at fixed impact parameter.   

The  measurements at the LHC energy of $\sqrt{s}=13$ TeV  indicates \cite{tamas} that the hadron  interaction region responsible for the inelastic processes is transforming from a black disk   to a black ring  with a reflective area in  the inner area of a ring. 
Transition to the black ring with the energy increase makes  the quantum fluctuations  a dominant source contributing to the observable event--by--event  multiplicity fluctuations. This process  is consistent with a general principle of  correlations relaxation.  It is due to the shrinkage of the range available for the impact parameter variations \cite{mdij}. 

The multiple production under hadron collision is a process described by the distribution over number $n$ of produced particles at fixed energy value 
\begin{equation}\label{pn}
P_n(s)\equiv \sigma_n(s)/\sigma_{inel}(s).
\end{equation}
The function 
\begin{equation}\label{pnb}
P_n(s,b)\equiv \sigma_n(s,b)/\sigma_{inel}(s,b)
\end{equation}
is the respective distribution at a particular value of the impact parameter of the colliding hadrons.\footnote{The functions are evidently normalized $\sum_nP_n(s)=1$, $\sum_nP_n(s,b)=1$.}
The inelastic overlap function has a prominent maximum at the impact parameter value of $b=R(s)$ when $s\to\infty$, $R(s)\simeq\ln s$ in this limit.
 It results in the following relation \cite{mdij} between the 
 introduced distributions, Eqs. (\ref{pn} and (\ref{pnb}), at $s\to\infty$: 
 \begin{equation}\label{flu}
 P_n(s)\simeq P_n(s,b)|_{b=R(s)}.
 \end{equation}
Fluctuations of multiplicity near the average value $\langle n\rangle (s)$ are happening due to the two above indicated sources.
The multiplicity fluctuations have quantum origin at $s\to\infty$ and are not related to the impact parameter variations. It happens  due to the reflective scattering mode leading to the black ring formation. Shrinkage of the effective $b$--range is a characteristic feature of this mode. The absorptive scattering mode, leading to the black disc scattering picture  at $s\to\infty$, does not assume such a conclusion.

Thus, the multiplicity distribution  $P_n (s)$  receives  contributions from the two sources of  a different origin at finite energies. Those are the fluctuations of multiplicity due to experimentally uncontrolled  $b$-dependence of the probability  
$P_n (s,b)$    and  quantum fluctuations of $n$ at fixed values of $b$. The possibility of their separation will be discussed in the next section.

The probability $P_n(s,b)$ enters into calculations  of a final state entropy and other thermodynamic quantities in hadron interactions. 

\section{Separating quantum fluctuations of multiplicity}
Now we are going to discuss possibilities of  experimental detection of the quantum fluctuations of multiplicity.

Quantum fluctuations of multiplicity are the  fluctuations  under  fixed values of $s$ and $b$ and they originate from a probabalistic nature of the wave function of colliding protons and dynamics of their interaction. Discussion  of the quantum fluctuations impact on  the observables can be found in \cite{mant}.  The perturbative QCD aspects of multiplicity fluctuation studies in hard processes and parton rescattering, in particular, have  recently been discussed in \cite{doc}. 

Direct way to assess  quantum fluctuations of multiplicity is to extract  the function $P_n(s,b)$ from  measurements of the multiplicity distribution $P_n(s)$ at fixed values of centrality. However,  experimentally determined centrality, $c_{\exp}$, has no one-to-one correspondence   with centrality $c_b$ \cite{centr} determined  by  the impact parameter value and therefore the $c_b$ needs to be somehow reconstructed from $c_{\exp}$. For proton--proton scattering, this could be similar to the case of proton--nucleus scattering\cite{pepin}.

To avoid such a reconstruction of the impact parameter value, we assume an absence of correlations between  the two types of multiplicity fluctuations, i.e. that one can represent this quantity  as the sum:
\begin{equation}\label{sum}
\delta n\equiv n-\langle n \rangle=\delta_b n+\delta_q n,
\end{equation}
where $\delta_b n$ are the fluctuations resulting from  the impact parameter variations and 
$\delta_q n$ is the contribution of the quantum fluctuations. 

The reflective scattering mode results in the limit  $\delta_b n\to 0$  at $s\to\infty$ \cite{mdij} and does not affect the quantum fluctuations of multiplicity. Asymptotically, the observed fluctuations should be associated with the quantum fluctuations, $\delta n=\delta_q n$.

How to separate quantum fluctuations of multiplicity at finite energies and possibly to measure them? To answer this question  it is instrumental using the above separation, Eq.(\ref{sum}), for the multiplicity fluctuations and the mechanism of multiparticle production proposed by Chou and Yang long time ago \cite{cymd}. Those authors proposed to relate experimentally observed  broad multiplicity distribution  to {\it ``an incoherent superposition of collisions at different impact parameters, each of which gives a narrow multiplicity distribution''}. It implicitly uses a classical aspect of the impact parameter as it was mentioned above. Thus, assuming separation of various contributions and 
since forward and backward multiplicities $n^b_F$ and $n^b_B$ share the same impact parameter value in the collision event, they should be  equal under this consideration.
Hence, according to \cite{cymd}, both values $n^b_F$ and $n^b_B$  can fluctuate, but those fluctuations are strongly interrelated. Such mechanism is valid for  both scattering modes, absorptive and reflective ones, but the range of multiplicity fluctuations  ``due to the impact parameter'' is shrinking  because of the reflective scattering mode is reducing the available  $b$--range with the energy increase \cite{mdij}.

In general, there is no reason to expect an existence of correlations in {\it quantum fluctuations} of multiplicities in the forward and backward hemispheres and their stochastic nature seems to be a natural assumption.  We do not concern here  an interesting problem of the quantum entanglement observation   at colliders and  reconciliation of this quantum phenomena with a local realism adopted in the experimental measurements (see for discussions e.g. \cite{low} and references therein). In this regard, we would just like  to mention the results of measurements of spin correlations at the LHC \cite{cms, atl} the origin of which could follow e.g. from the total angular momentum conservation \cite{cymd} and   the reflective scattering mode  (antishadowing) effect for the spin correlation parameter in the top quark production at the LHC  \cite{pl,mpla}.

Observation of a nonvanishing difference $\Delta \equiv n_F-n_B$ would  reveal  presence of
 the quantum fluctuations of multiplicity, which  are presumably stochastic.

The distribution $P_{\Delta }(s)$\footnote{The distribution is an even function in the variable $\Delta $ in $pp$--collisions, i.e. it is invariant under replacement $\Delta \to -\Delta $ and is, in fact, function of $\Delta^2$.} should be sensitive to the properties of quantum fluctuations of multiplicity and its experimental studies would provide knowledge on the origin, magnitude and other properties of quantum fluctuations.

Analogy with quantum optics   can be useful for modeling   the particle distributions at   finite energies. The
use of 
 gamma and negative binomial distributions was shown to be a
 relevant option  for nuclei--nuclei and hadron--nuclei reactions \cite{rog}.

 Extension of this,   nuclear based approach, to description of the small systems such as hadrons and their interactions is supported by  experimentally discovered similarities in the  behavior of observables in both types of reactions, e.g. discovered ridge and other collective effects observed under interactions of small systems.     Application to  hadron collisions is complimentary gaining advantage from   validity of   unitarity condition.  It allows to relate centrality $c_b$ with cumulative activities in  hadron case \cite{centr}.

Since the above distributions are of a nonstochastic nature, it is difficult to expect their relevance  for the asymptotic energy region where dominance of a quantum multiplicity fluctuations takes place. These distributions  incorporate  combined effect of  the both types of fluctuations  and should be associated therefore with the available energies. One can expect evolution with energy towards better agreement of $P_{\Delta }(s)$   with  Poisson or binomial distribution when the collision energy increases.
\section*{Conclusion}
It is proposed to study\footnote{Similar measurements have been performed by  UA5 collaboration at CERN S$\bar p p$S \cite{ua5}. } distribution of the difference between the forward and backward multiplicities, $P_{\Delta} (s)$, $\Delta\equiv n_F-n_B$,\footnote{The multiplicities $n_F$ and $n_B$  should correspond to  the same event.}  to detect quanum fluctuations of the total multiplicity.   Presence of quantum fluctuations could be detected by nonvanishing distribution function $P_\Delta$. 

The reason for its nonvanishing   value  could  stochastic nature of the quantum fluctuations of multiplicity.  Effects of quantum entanglement which generate correlations could compensate these multiplicity fluctuations $P_{\Delta}$, and magnitude of $P_{\Delta}$ at $\Delta=0$ could be used for estimation of  their randomness.    Form of a distribution over  $\Delta $ would be interesting for studying dynamics of hadron interactions, origin of quantum fluctuations, presence of  their correlations, and  possible connection with the initial hadron state  together with properties of the hadron interactions and transient state.

For receiving an additional  information and to obtain more definite results, the above measurements could be combined with the multiplicity distribution $P_n(s)$ measurements performed at fixed values of centrality.

Importance of the proposed measurements is rather general for common and independent studies of  multiparticle dynamics. Studies of quantum fluctuations of multiplicity at colliders could include symmetrical collisions of nucleai.

\end{document}